# Built-in Bernal gap in large-angle-twisted monolayer-bilayer graphene

Alex Boschi[1], Zewdu M. Gebeyehu[1,2], Sergey Slizovskiy[3,4], Vaidotas Mišeikis[1,2], Stiven Forti[1,2], Antonio Rossi[1,2], Kenji Watanabe[6], Takashi Taniguchi[7], Fabio Beltram[8], Vladimir I. Fal'ko[3,4,5], Camilla Coletti[1,2,*], Sergio Pezzini[8,**]

[1]*Center for Nanotechnology Innovation @NEST, Istituto Italiano di Tecnologia, Piazza San Silvestro 12, I-56127 Pisa, Italy*
[2]*Graphene Labs, Istituto Italiano di Tecnologia, Via Morego 30, 16163 Genova, Italy*
[3]*National Graphene Institute, The University of Manchester, Manchester, M13 9PL, UK*
[4]*School of Physics & Astronomy, The University of Manchester, Oxford Rd., Manchester, M13 9PL, UK*
[5]*Henry Royce Institute for Advanced Materials, Manchester, M13 9PL, UK*
[6]*Research Center for Electronic and Optical Materials, National Institute for Materials Science, 1-1 Namiki, Tsukuba, 305-0044, Japan*
[7]*Research Center for Materials Nanoarchitectonics, National Institute for Materials Science, 1-1 Namiki, Tsukuba, 305-0044, Japan*
[8]*NEST, Istituto Nanoscienze-CNR and Scuola Normale Superiore, Piazza San Silvestro 12, I-56127 Pisa, Italy*

* camilla.coletti@iit.it

** sergio.pezzini@nano.cnr.it

**Abstract**

Atomically thin materials offer multiple opportunities for layer-by-layer control of their electronic properties. While monolayer graphene (MLG) is a zero-gap system, Bernal-stacked bilayer graphene (BLG) acquires a finite band gap when the symmetry between the layers' potential energy is broken, usually, via a displacement electric field applied in double-gate devices. Here, we introduce a twistronic stack comprising both MLG and BLG, synthesized via chemical vapor deposition, showing a Bernal gap in the absence of external fields. Although a large (~30°) twist angle decouples the MLG and BLG electronic bands near Fermi level, proximity-induced energy shifts in the outermost layers result in a built-in asymmetry, which requires a displacement field of 0.14 V/nm to be compensated. The latter corresponds to a ~10 meV intrinsic BLG gap, a value confirmed by our thermal-activation measurements. The present results highlight the role of structural asymmetry and encapsulating environment, expanding the engineering toolbox for monolithically-grown graphene multilayers.

**Introduction**

The electronic properties of few-layer graphenes are sensitive to layer number[1], stacking order[2,3], lack or presence of inversion symmetry [4,5], external electromagnetic fields[6], as well as to the twist angle between consecutive layers (their relative crystallographic orientations)[7]. Based on this last tuning parameter, the band hybridization between contiguous sheets can be artificially modulated, resulting in spectacular emergent behaviors for small-angle values associated to specific periodicities of the interlayer moiré[8]. This paradigm, initially demonstrated in twisted bilayer graphene (TBG) [9,10], has now been extended to a plethora of related moiré systems [11], including twisted MLG-BLG (TMBG). For twist angles ~1°, TMBG has been shown to support flat bands promoting correlated and topological phases, including insulators [12,13], orbital magnets [14–16] and charge density waves [17]. The lack of inversion symmetry makes the electronic response of small-angle TMBG strongly dependent on the direction of an external displacement field (*D*) that can be controlled by top and bottom gate electrodes [12–15,17]. In the limit of large twist angle (where moiré effects are negligible), TMBG is expected to decompose into two low-energy subsystems that retain unperturbed MLG and BLG character, mimicking the independent-MLGs behavior found in large-angle TBG [18–23]. However, a complete picture of large-angle TMBG needs to account for its structural asymmetry, which can have a subtle (yet measurable) influence on the exact band structure close to Fermi level. Specifically, not only BLG is sensitive to *D* – which drives band-gap opening [24–26] exploitable for quantum confinement [27–29] and multiple phase transitions [30–32] – but also, when embedded in stacks, to environment-induced interlayer asymmetry [5,24], as demonstrated by the observation of a mini-gap in large-angle-twisted double BLG (TDBG) [33]. Building on these findings, Agarwal *et al.* recently reported optoelectronic properties widely beyond the independent-subsystems scenario of TDBG [34]. These results motivate further exploration of BLG-containing stacks – such as large-angle TMBG – as well as the development of scalable synthesis methods for these systems – such as chemical vapor deposition (CVD) growth of single crystals [35].

In this work, we provide experimental evidence (supported by detailed mesoscale modelling) for a built-in band gap in a BLG studied as a part of an hBN-encapsulated large-angle TMBG. Key to our observations is the ability to directly grow monolithic TMBG crystals via low-pressure CVD (LP-CVD) and integrate them into high-quality van der Waals (vdW) devices.

## Results and Discussion

A side-view sketch of the studied devices is shown in Figure 1a. We employ a dual-gate configuration to perform transport measurements with independent control of gate-induced charge density ($n_{tot}$) and *D* (defined in Methods; optical microscopy images of the two TMBG devices investigated are shown in Supplementary Note 1, Figure S1). The trilayer graphene structure, encapsulated in hexagonal boron nitride (hBN), can be separated into a MLG (blue) rotated by ~30° with respect to two aligned layers forming a BLG (red and orange). In analogy with 30°-TBG [21], we expect the lattices of the two subsystems to arrange into an incommensurate configuration lacking translational symmetry (i.e., with no moiré superlattice), as sketched in Figure 1b. The real-space twist leads to a large momentum mismatch between the MLG and BLG bands located at the Brillouin zone corners (see Figure 1c), effectively leading to their decoupling at low energy. Based on previous results for 30°-TBG, the features of real-space quasicrystalline registry could appear only at high energies (few eV) [36,37], inaccessible in solid-state transport devices. The TMBG crystals are synthesized via the LP-CVD process introduced for 30°-TBG in Ref. [21], in which the graphene-Cu interaction locks the possible interlayer twist angles to either 0° or 30° [38], with increased growth times (here, 60 min) favoring the formation of concentric multilayer structures, sharing a single nucleation point (see Methods for experimental details). In these samples, the hexagonal shape of the layers reflects their relative crystallographic orientation, making TMBG easily recognizable after transfer to $SiO_2$/Si substrates (see Figure 1d) among other crystals with varied 0° or 30° rotation and layer number. Our growth approach replaces the tear-and-stack procedure ubiquitously employed for twisted graphene devices [39,40], including small-angle TMBG [12–16] and large-angle TDBG [33,34], highlighting a possible path towards scalability of this class of materials.

A first experimental indication of the composite MLG-BLG band structure of our TMBG [41] is obtained from the 2D Raman mode, which couples to the electronic degrees of freedom in graphene multilayers [42]. As shown in Figure 1e (black continuous line), we observe a narrow peak centered at ~2690 $cm^{-1}$ with two asymmetric shoulders, suggesting the convolution of a single-Lorentzian peak (resulting from the Dirac dispersion of MLG) with a broad multi-Lorentzian peak (resulting from the four-band parabolic dispersion of BLG). The measured TMBG 2D mode can be quantitatively reconstructed as the sum (grey continuous line) of two Raman spectra separately acquired on 30°-TBG (blue dotted line) and BLG (grey dotted line, multiplied by a factor two) crystals from the same growth batch (full spectra are shown in Supplementary Note 2, Figure S2). We recall that the 2D peak of 30°-TBG is indistinguishable from that of MLG, apart from a blueshift attributable to the modified dielectric environment [21], which also applies to the MLG subsystem within TMBG.

Before discussing the results of electrical transport experiments, we discuss the expected spectral properties of electrically-biased TMBG, at zero and quantizing magnetic fields. This analysis is based on a self-consistent solution for charge and potential distribution in vertically biased few-layer graphene, taking into account the out-of-plane dielectric polarizability of each layer, using a model developed and tested in Ref. [22]. For TMBG in quantizing magnetic fields, the charge distribution is determined by the Landau-level (LL) pinning between its monolayer and bilayer components, stemming from the quantum-capacitance effect [22,43] (see Supplementary Note 3 and Figure S3 for details). In this analysis we take into account proximity-induced energy shifts (δ) on the graphene layers interfacing encapsulating hBN, relative to the on-layer electrostatic potential of the middle one (interfacing carbon layers on both sides). Such energy shifts are in line with the recent model developed by Tseng and Chou for free-standing (i.e., interfaced with vacuum) small-angle TMBG [44]. While the actual magnitude of δ is not known (would depend on the encapsulating material), we find that it can be determined from the comparison of the computed (in)compressibility maps, Figs. 1g & 1h, with experimentally measured quantum Hall (QH) maps in Fig. 2e. In Figs. 1g & 1h we plot the density of states (DOS) of TMBG as a function of $n_{tot}$ and $D$ for the cases of $\delta = 0$ and $\delta = 18$ meV, respectively. The color-coding is chosen to indicate when the chemical potential of electrons is pinned at the LLs (bright color) and when it crosses the inter-LL gaps (dark color). Diagonal (vertical) lines trace MLG (BLG) LLs, and their crossings correspond to mutual pinning of MLG and BLG LLs. The simulation of DOS maps was iterated for different values of δ, until best matching the experimental QH data.

The influence of proximity shift, δ, can be appreciated from the relative shift of compressibility maps in Figs. 1g & 1h. Its influence is also illustrated in Figure 1f, where we plot the magnitude of the band gap in the BLG subsystem ($\Delta$) as a function of $D$, self-consistently calculated for δ = 0 (black line) and δ = 18 meV (dark cyan line). In particular, the proximity shifts induce a finite $\Delta$ even in the absence of an external $D$-field. Furthermore, for each value of δ, one can identify a displacement field, $D_c$ (see inset in Fig. 1f), that compensates the proximitized interlayer asymmetry and closes the BLG gap, suggesting a practical tool for precise measurements of the energy shift on graphene under different encapsulating environments.

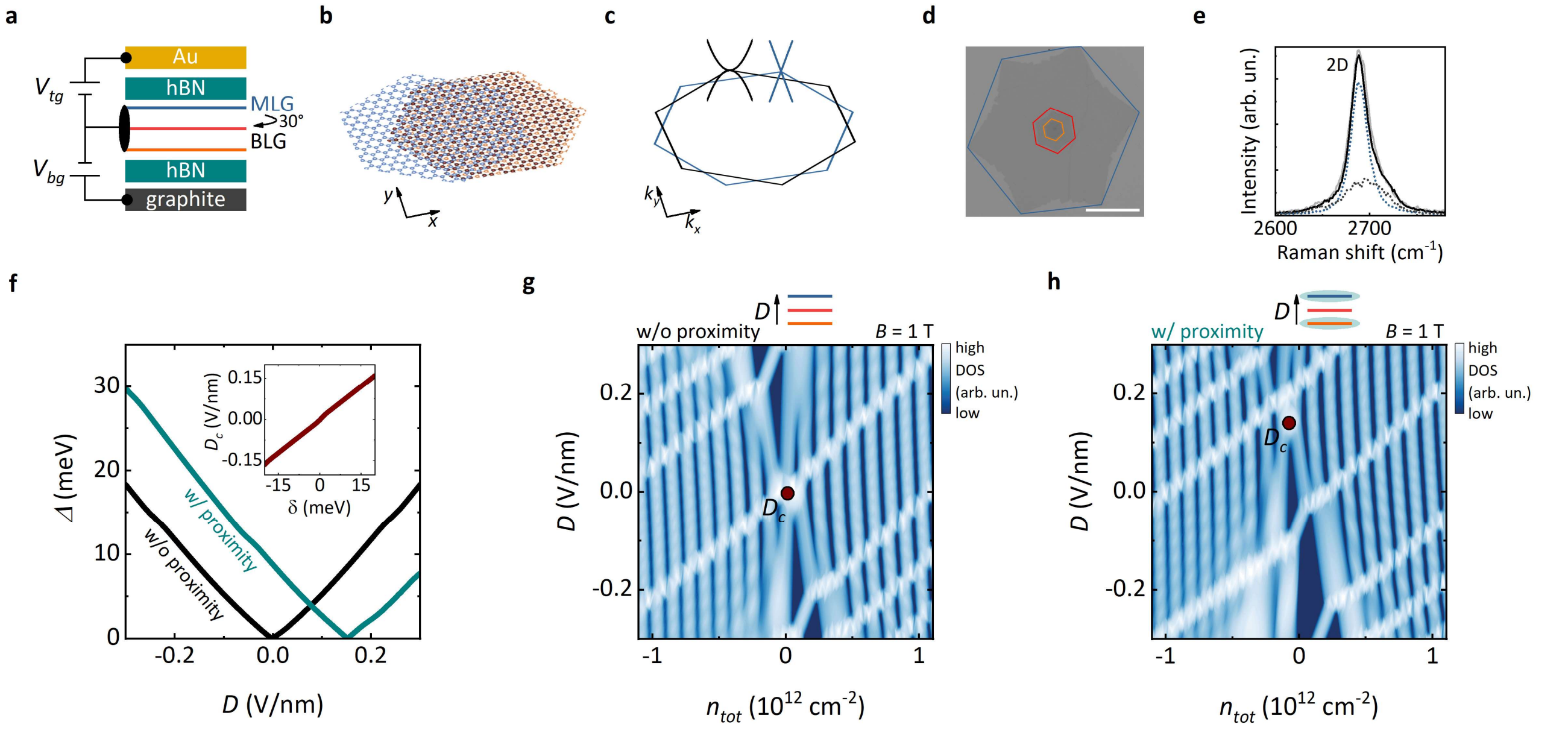

**Figure 1: Device concept, CVD-grown twisted graphene crystals and effect of proximity energy shifts .** *(a) Side-view sketch of the studied devices. The gating scheme is indicated on the lefthand side. The three graphene layers are represented, from top to bottom, as a blue, red and orange line. (b) Real-space arrangement of the TMBG crystals: MLG (blue lattice) is superimposed to BLG (red and orange lattices, following the color scheme of panel (a)) with a 30° twist angle. (c) Sketch of the reciprocal-space configuration, with the MLG (blue) and BLG (black) Brillouin zones and low-energy bands at K. (d) Optical microscopy image of a representative CVD-grown trilayer graphene crystal transferred on $SiO_2$/Si. The three hexagons mark the concentric graphene layers forming TMBG (the color scheme follows panel (a)). The scale bar is 40 µm. (e) 2D Raman mode measured on TMBG on $SiO_2$/Si (black). The grey continuous curve is the sum of a MLG-like (dotted blue line) and a BLG (dotted grey line) component. (f) BLG band gap as a function of displacement field, without including proximity energy shifts (δ = 0, black line), and considering the shifts (δ = 18 meV, dark cyan line). Inset: gap-closing displacement field as a function of δ. (g,h) DOS of TMBG at B = 1 T calculated without including proximity energy shifts (δ = 0) (g) and considering the shifts*

*(δ = 18 meV) (h). The dark red circle indicates the BLG gap closing. The sketch above each graph shows the TMBG layers (following the color scheme of panel (a)), the arrow indicates the direction of positive D-field. Dark cyan shadows highlight the graphene layers interfacing encapsulating hBN, hence affected by δ.*

In Figure 2 we present gate-dependent electrical-transport measurements performed at $T$ = 0.36 K on TMBG device D1 (analogous data from D2 are shown in Supplementary Note 4, Figure S4). Figure 2a shows the longitudinal resistivity $\rho_{xx}$ as a function of the top and bottom gate bias ($V_{tg}$ and $V_{bg}$). At large carrier density (top right and bottom left corners), $\rho_{xx}$ exhibits low values (<20 Ω) that indicate high carrier mobility, as typically observed in hBN-encapsulated graphene stacks. At low carrier density, we observe two intersecting features, with a region of high resistivity ($\rho_{xx}$ up to ~10 kΩ) departing from $V_{tg}$ = $V_{bg}$ = 0 and extending only in the lower-right quadrant. To better discern the origin of the different transport contributions, in Figure 2b we show $\rho_{xx}$ data acquired within the dotted rectangle in Figure 2a, as a function $n_{tot}$ and $D$. Two split features, attributable to the charge neutrality points (CNPs) for the two subsystems, are visible. The first one, dispersing almost vertically along $n_{tot}$ = 0, is associated to the subsystem with larger DOS, i.e., BLG. The second one, following a diagonal trajectory due to screening of the gate potentials by BLG, tracks the MLG CNP. We observe a marked slope change of the MLG CNP in the vicinity of $D$ = 0 – signaling decreased screening and hence suppressed DOS in BLG – accompanied by large resistivity. At large negative $D$, the MLG CNP reverts to the original slope and, at the same time, resistivity is suppressed, indicating restored DOS in the BLG subsystem. This behavior is fully consistent with the MLG CNP crossing a band gap in BLG [33] and can be reproduced by calculations shown in Figure 2c, inset (see Supplementary Note 3 for details). Overall, we notice the pronounced asymmetry with respect to the displacement field which reflects the structural asymmetry of TMBG (as opposed to measurements on structurally symmetric 30°-twisted MLG-BLG-MLG, shown in Supplementary Note 5, Figure S5). This characteristic is highlighted by $\rho_{xx}$ curves at selected $D$ values, shown in Figure 2c. At $D$ = 0 we observe a clear resistivity peak, which increases (and slightly shifts) at $D$ = -0.1 V/nm; upon reversing displacement field ($D$ = +0.1 V/nm), the resistivity peak disappears.

We then apply a moderate magnetic field ($B$ = 1 T, Figure 2d-f) perpendicular to the TMBG device, that reveals the independent Landau quantization of the two subsystems. In Figure 2d, the longitudinal conductivity $\sigma_{xx}$ (obtained by combining longitudinal and Hall resistivity data via the classical tensorial relation) shows two separated sets of oscillations, with different slope and frequency. The low-frequency oscillations are associated to the subsystem with smaller DOS, i.e., to MLG. The MLG oscillations show an uneven spacing, reflecting the energy separations between the LLs of massless Dirac Fermions, as previously observed in capacitively-coupled graphene layers, either spaced by dielectrics [45] or twisted by a large angle [23]. The high-frequency oscillations are associated to BLG and are absent in an extended region showing vanishing $\sigma_{xx}$ (dark blue in Figure 2d), which signals the BLG gap. Figure 2e ($\sigma_{xx}$ as a function of $n_{tot}$ and $D$) clearly shows that the gapped region collapses under a displacement field $D_c$ = 0.14 V/nm (and reopens for $D > D_c$), which compensates the built-in BLG asymmetry. The measured pattern of LLs is best matched by DOS calculations shown in Figure 1h, which include an energy shift δ = 18 meV. The sign of $D_c$ indicates that the intrinsic BLG polarization points from top to bottom, in accordance with Ref. [44]. Based on the magnitude of $D_c$, we estimate a corresponding BLG band gap $\Delta$ ~ 10 meV, comparable to the one reported for TDBG [33,34]. Although spontaneous, few-meV gaps due to strong electron-electron interactions were measured in suspended BLG devices [46–48], those gaps show distinctive $D$-symmetric gap closing and reopening, ruling out a possible relation with the observed phenomenology. When the MLG zero-energy LL (0-LL) crosses the BLG gap, we clearly detect MLG-like QH states at filling factor $\nu_{tot} = \pm 2$, as shown by $\sigma_{xx}$ and $\sigma_{xy}$ curves in Figure 2f. Outside of the BLG gap, we observe quantized steps in $\sigma_{xy}$ with the expected amplitude 4e$^2$/h (due to spin and valley degeneracy), with 8e$^2$/h steps in case of coincidence of LLs from the two subsystems.

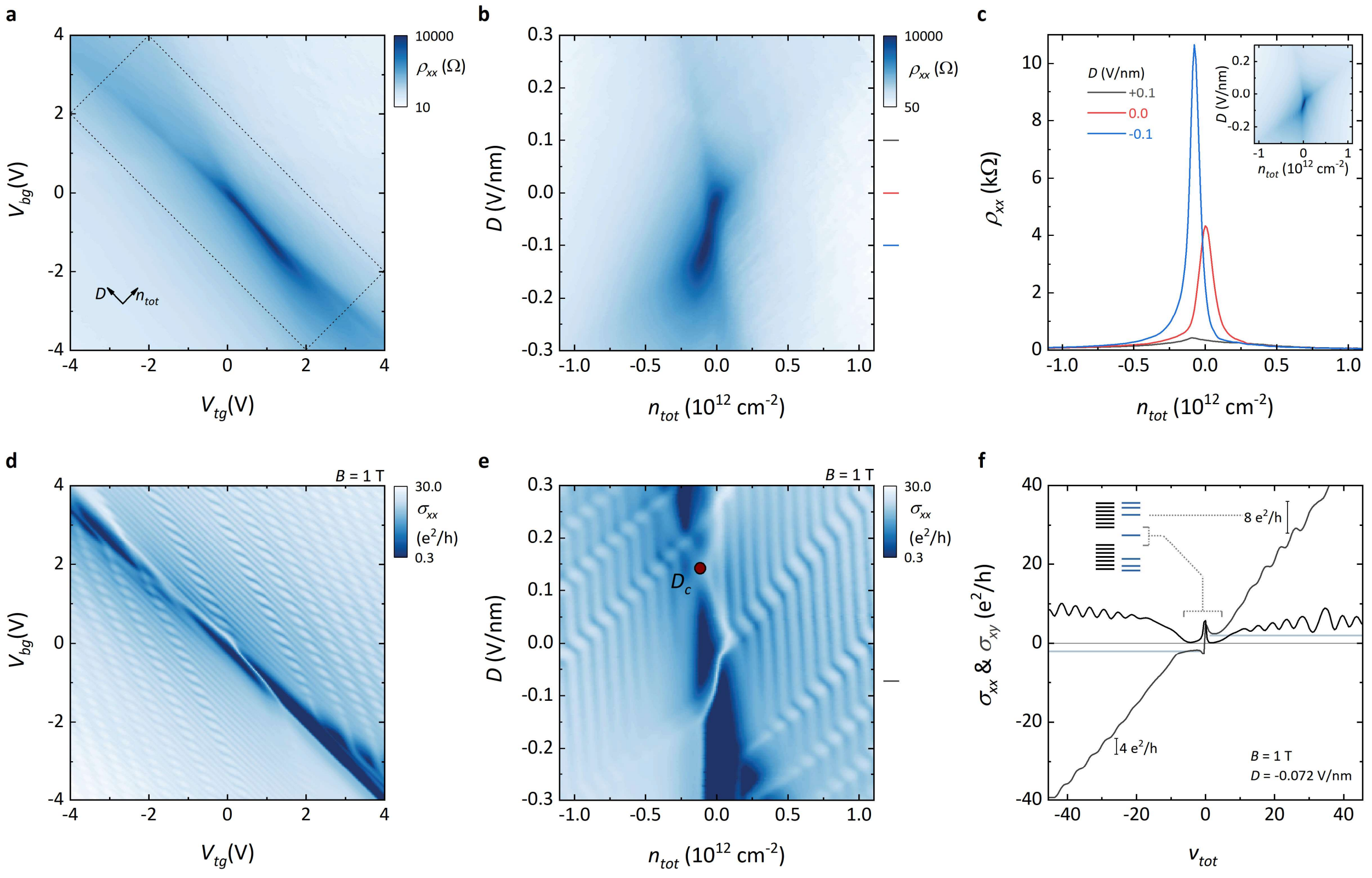

**Figure 2: Low-temperature (magneto)transport of large-angle-twisted monolayer-bilayer graphene.** *(a) Longitudinal resistivity as a function of the gate potentials. (b) Longitudinal resistivity as a function of carrier density and displacement field, acquired within the dotted rectangle in panel (a). (c) Resistivity as a function of carrier density at selected displacement fields, as marked in (b). Inset: calculated resistivity of TMBG, as a function of carrier density and displacement field (the color scale is the same used in panel (b)). (d) Longitudinal conductivity at B = 1 T as a function of the gate potentials (same ranges*

*as in (a)). (e) Longitudinal conductivity at B = 1 T, as a function of carrier density and displacement field (same ranges as in (b)). The dark red circle indicates the position of gap closing in BLG, which identifies the displacement field $D_c$ compensating the intrinsic BLG polarization. (f) Longitudinal (black) and Hall (grey) conductivity as a function of the total filling factor $\nu_{tot} = n_{tot}h/eB$ at D = -0.072V/nm at B = 1T (see marker in panel (e)). The inset shows a sketch of the alignment of LLs from the two subsystems (blue for MLG, black for BLG). The MLG 0-LL lies within the BLG gap, leading to a MLG-like ±2 $e^2/h$ QH sequence at small fillings. 8 $e^2/h$ steps in the Hall conductivity signal coincidence of LLs from the two subsystems (4 $e^2/h$ steps are measured otherwise). All measurements are performed at T = 0.36 K on device D1.* A logarithmic color scale is used in panels (a), (b), (d), (e).

Having identified the experimental signatures expected for an intrinsic ~10 meV Bernal gap in dual-gated TMBG, we now proceed to its quantitative estimate. In Figure 3a, we present isotherms of $\rho_{xx}$ as a function of $n_{tot}$, measured at $D$ = -0.075 V/nm on device D1 (analogous data from D2 are presented in Supplementary Note 6, Figure S6). We observe thermal activation of the resistivity for $T$ < 100 K, confirming the presence of an energy gap. This behavior is consistently observed at different $D$ values, within the previously identified high-resistivity region (where the MLG CNP lies within the BLG gap). To appreciate changes in the conductivity of the BLG subsystem as a function of temperature, we minimize the conductivity contribution of the other subsystem, i.e. MLG, by considering the resistance maximum corresponding to MLG CNP. As shown in Figure 3b, we find an Arrhenius-type dependence for the maximum of the resistivity peak $\rho_{xx}^{max} \propto \exp[\Delta/(2k_BT)]$ (where $\Delta$ is the band gap and $k_B$ is the Boltzmann constant). The extracted gap reaches a maximum of $\Delta$ = 15 meV at $D$ = -0.075 V/nm, in line with calculations shown in Figure 1f for δ = 18 meV. The $D$-dependence shown Figure 3b inset is influenced by MLG CNP crossing the BLG gap as a function of the gate voltages (see trajectory in Supplementary Figure S3g), since the thermal activation gap reflects the distance between MLG CNP and the BLG band edges. As the displacement field becomes increasingly negative, the BLG gap widens (see Figure 1f, dark cyan line), however MLG CNP approaches one of the BLG band edges, leading to a lower carrier activation energy and thus to a decrease of measured Δ. The band gap measured from thermal activation is finite even at $D$ = 0, confirming that no external electric field is required for BLG gap opening within TMBG.

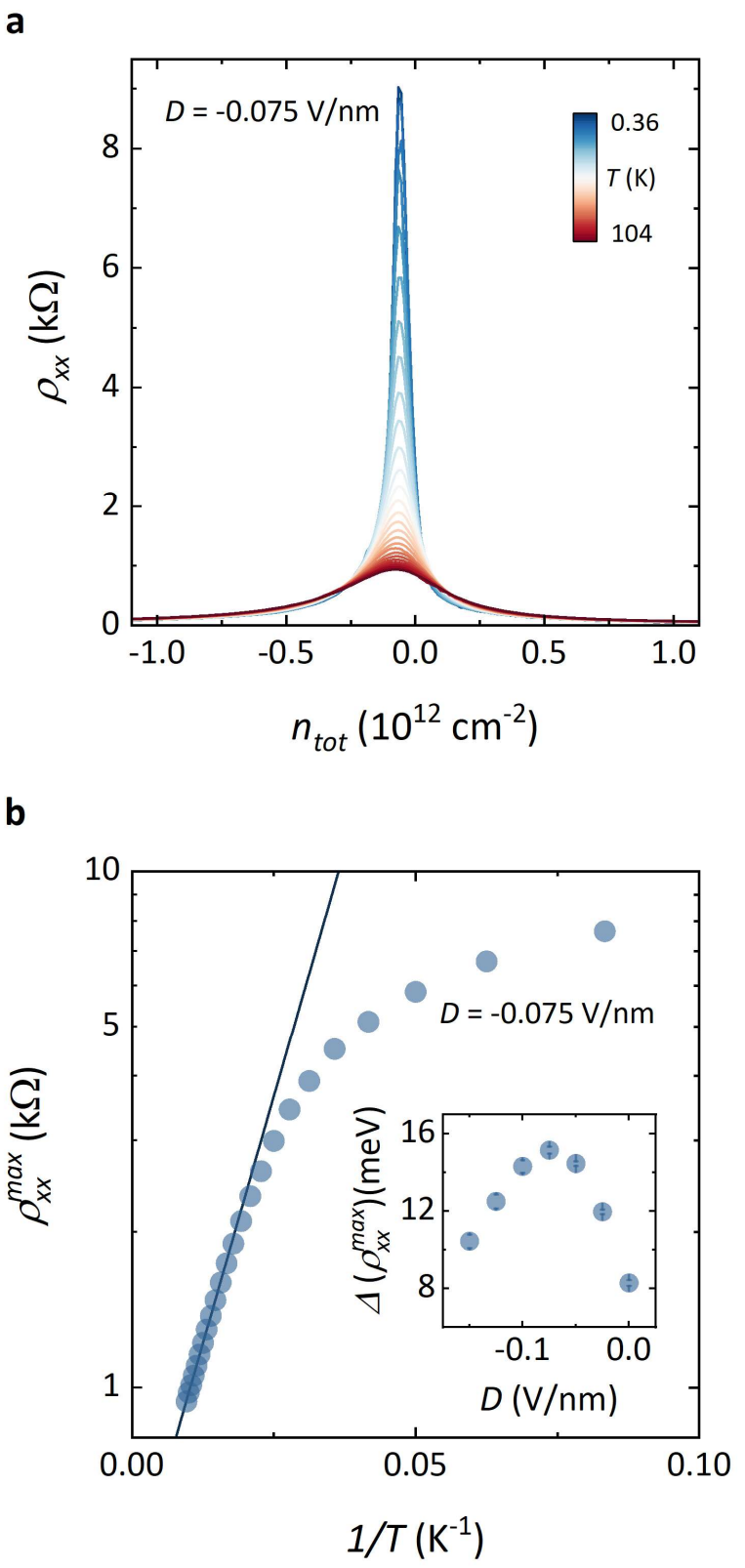

**Figure 3: Thermal activation across the Bernal gap**. *(a) Longitudinal resistivity isotherms as a function of the carrier density, measured on sample D1 at D = -0.075 V/nm for 0.36 K < T < 104 K. (b) Arrhenius plot of the resistivity peak maxima from data in (a) (blue dots). The fit to* $\rho_{xx}^{max} \propto \exp[\Delta/(2k_BT)]$ *is shown as a blue continuous line. Inset: activation gaps obtained from the Arrhenius fit at different displacement fields. Error bars correspond to ± one standard error from the fits.*

We further investigate the transport properties of TMBG under large perpendicular magnetic fields, which allows us to investigate the evolution of the multicomponent LLs of the two subsystems promoted by Coulomb interactions [49]. In Figure 4a and 4b, we show $\sigma_{xx}$ and $\sigma_{xy}$ data at $B$ = 4 T, as a function of $\nu_{tot}$ and $D$ (the measurements span the usual $n_{tot}$ and $D$ ranges). For the MLG subsystem we observe only the 0-LL, which splits into four branches (bounding QH gaps at $\nu_{MLG}$ = 0, +1, -1) due to complete lifting of the spin and valley degeneracy. The trajectories of the four MLG components are modulated by multiple crossings with LLs from the BLG subsystem, which also clearly show degeneracy lifting. In particular, the BLG 0-LL possesses an extra two-fold degeneracy associated with the orbital degree of freedom, making the multicomponent QH effect even richer and $D$-tunable [50]. We observe a clear symmetry of the BLG pattern with respect to the gap closing point at $D = D_c$, further confirming the intrinsic BLG polarization and its compensation by a positive displacement field. At $B$ = 8 T (Figure 4c and 4d) we observe multiple $D$-driven phase transitions in the BLG 0-LL around $D = D_c$, which reproduce recent results on BLG samples around $D = 0$ [51]. These observations indicate that: (i) the quality of our CVD-based devices is fully comparable with samples based on mechanically exfoliated graphene layers; (ii) once the intrinsic polarization is compensated, the physics of the BLG subsystem embedded in TMBG is essentially indistinguishable from that of a stand-alone BLG. In Figure 4e and 4f we show $\sigma_{xx}$ and $\sigma_{xy}$ curves at $B = 8$ T and selected values of displacement field. Figure 4e shows the full broken symmetry of the MLG 0-LL within the BLG band gap, while Figure 4f highlights the full broken symmetry in of the BLG 0-LL (note that in this case the plateau sequence is shifted by a factor $-2e^2/h$ due to the contribution of two parallel-conducting edge channels from hole-doped MLG).

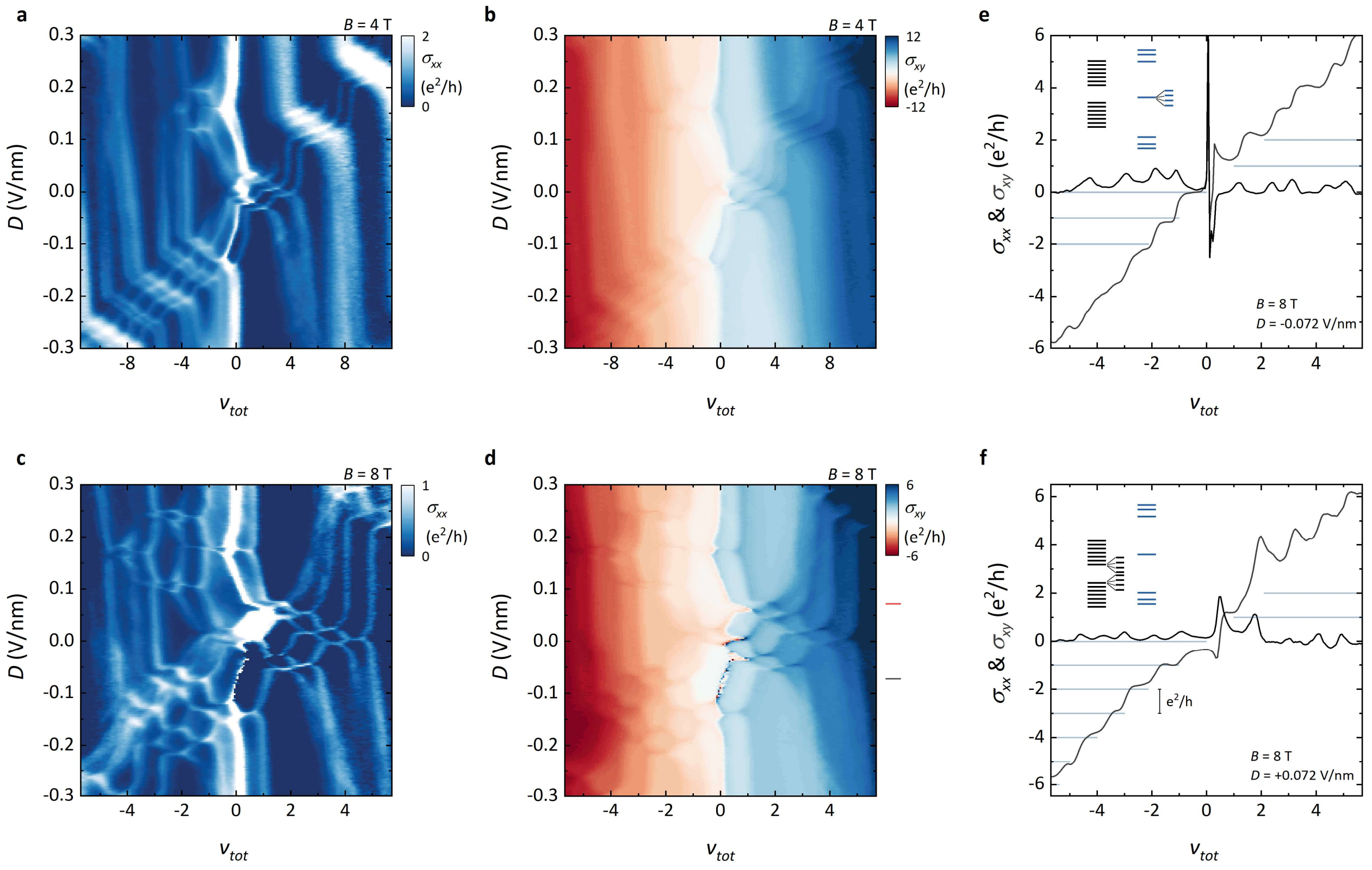

**Figure 4: Multicomponent Landau levels in the two subsystems.** *Longitudinal and Hall conductivity measured on device D1 at B = 4 T (a,b) and B = 8 T (c,d). Curves at selected displacement field (indicated by the grey and red marks in panel (d)) are shown in (e) and (f), respectively; insets: sketches of the alignment of LLs from the two subsystems (blue for MLG, black for BLG), with degeneracy lifting in MLG (e) and BLG (f). Data acquired at T = 0.36 K on device D1.*

## Conclusions

In conclusion, we established the presence of a built-in BLG band gap in large-angle TMBG. The gap is induced by the lack of inversion symmetry, combined with proximity energy shifts on the outmost graphene layers. The BLG gap leads to specific signatures in electrical-transport properties, both under zero and quantizing magnetic fields that can be employed for its quantitative estimate. Analogous features have also been reported in a recent pre-print [52], where TMBG was obtained from exfoliated graphene flakes. Although our samples are encapsulated in hBN (as now ubiquitous in graphene research), we note that the observed effects are expected to occur also in free-standing TMBG, where even larger gaps are predicted [44], and in TMBG encapsulated with other dielectrics, which should modify the gap magnitude. Along this line, we can anticipate the use of TMBG and measurements of the BLG gap closing at $D_c$ in dual-gated devices as a metrology platform for determining proximity-induced energy shifts caused by various encapsulating materials or insulating substrates (such as transition metal dichalcogenides). Moreover, the built-in BLG gap could be leveraged for broadband detection in the THz window (following recent findings on TDBG [34]), with tunable absorption properties depending on the encapsulating environment. Finally, let us remind the high interest in BLG gap for the definition of quantum point contacts [27] and dots [28,29], which might be designed with approaches beyond the current paradigm of local gating. The applicative potential of TMBG-based (opto)electronic and quantum devices is reinforced by the large-scale growth method employed in the fabrication of devices used in the reported studies.

## Methods

*Gate induced carrier density and displacement field:*

We adopt the relations $n_{tot} = 1/e \times (C_{tg}V_{tg}+C_{bg}V_{bg})$ and $D/\varepsilon_0 = (C_{bg}V_{bg}-C_{tg}V_{tg})/2$ ,[18] where $C_{bg}$ and $C_{tg}$ are the capacitances per unit-area of the bottom and top gate, $V_{bg}$ and $V_{tg}$ are voltage bias applied to the bottom and top gate electrodes, $e$ is the electron charge, $\varepsilon_0$ is the vacuum dielectric permittivity. Both the top and bottom hBN crystals used in the devices are ~30 nm thick (as determined by atomic force microscopy, AFM). We employ $\varepsilon_r$ = 3 for the out-of plane dielectric constant of hBN [53,54], giving $C_{bg} = C_{tg} = 8.85 \times 10^{-8}$ Fcm$^{-2}$.

*CVD growth of multilayer graphene with 0°/30° twist:*

TMBG crystals are grown by chemical vapor deposition in a Aixtron Black Magic reactor (Aixtron 4” BM-Pro) on electropolished Cu foil at a pressure of 25 mbar and a temperature of 1065 °C. After electropolishing, the Cu foil is heated in air on a hot plate at 250 ˚C for 15 min to oxidize the Cu surface. The Cu foil is then loaded in the CVD reactor and annealed in Ar atmosphere at 1065 °C for 10 mins, followed by 60 minutes of graphene growth in a mixture of Ar, $H_2$ and $CH_4$ with a flow ratio of 900:40:0.8 sccm, respectively. Then the CVD reactor is cooled down to 100°C in an Ar atmosphere. After growth, the shape of graphene crystals shows a deviation from perfect hexagons due to diffusion-limited growth mechanisms, as typical for oxygen-rich Cu foil [55] and low partial pressure of $H_2$ [56] . Crystals obtained under these conditions show regular electron diffraction patterns [21] and low strain [57] (<1% after transfer to $SiO_2$/Si; though strain influences details of the BLG dispersion, the system is expected to remain gapless at such values [58]). We observe concentric multilayers with different interlayer rotations, each one locked to either 0° or 30° (with a preponderance of aligned layers, attesting at ~60-70%). Among the trilayer crystals, a ~30% fraction shows the stacking configuration targeted for this study.

*vdW assembly and device fabrication:*

Before vdW assembly, CVD-grown TMBG crystals are transferred to a $SiO_2$/Si substrate using a semi-dry technique involving electrochemical delamination with a PMMA/PPC carrier membrane [57]. We first prepare a crystalline bottom gate by picking-up a graphite flake (~5 nm thick) using an hBN flake carried by a PC/PDMS stamp [59]. The graphite back-gate screens disorder from the underlying $SiO_2$, yielding clean transport properties in gapped BLG devices [60]. After release and cleaning in chloroform, we use contact-mode AFM to mechanically clean the hBN surface [61]. A second hBN flake is used to pick-up a portion of the TMBG crystal. The hBN/TMBG is subsequently released at 180°C on top of the cleaned hBN/graphite. The devices are processed combining electron-beam lithography, reactive ion etching and thermal evaporation of Cr/Au edge contacts and top-gates.

*Raman spectroscopy:*

A commercial Renishaw “InVia” system is used for Raman characterization. Spectra are acquired employing a 532 nm wavelength and with a 100x objective, producing a fluence ~350 μW/ $\mu m^2$ on the sample, using a 1800 l/mm grating. Calibration of the system is performed using the Si Raman peak at 520 $cm^{-1}$.

*Electrical transport measurements:*

The electrical transport measurements are performed in a dry “ICE 300mK He-3 Continuos” cryostat equipped with an 8 T superconducting magnet (sample D1) and in a dry “ICE 3K INV” cryostat (sample D2). Four-probe measurements are performed with low-frequency (~13 Hz) lock-in detection, either in a constant current (~100 nA), or constant voltage configuration (0.1 mV). The source-drain current and longitudinal and Hall voltage drops are simultaneously recorded, while biasing the top and bottom gates using a dc source-meter.

**Data availability Statement**

The data presented in this study are available at https://doi.org/10.5281/zenodo.14178601.

**Author Information**

*Contributions*

S.P. conceived and designed the experiments. Z.M.G. and V.M. synthesized the TMBG crystals and transferred them to $SiO_2$/Si. A.B. and S.P. fabricated the devices and performed the transport measurements. A.B. and S.P performed the data analysis, with support from A.R. and S.F. S.S. and V.I.F. performed the theoretical modelling.

K.W. and T.T. provided hexagonal boron nitride crystals. F.B. and C.C. supported the experiments and coordinated the collaboration together with V.I.F and S.P. A.B. and S.P. co-wrote the manuscript with input from all co-authors.

*Competing Interests*

The authors declare no competing interests.

**Acknowledgments**

This work has received funding from: PNRR MUR project PE00000023 – NQSTI and the European Union's Horizon 2020 Research and Innovation Programme under Grant Agreement No. 881603 Graphene Flagship. We acknowledge funding from the European Union through the GraPh-X project (Grant agreement ID: 101070482). K.W. and T.T. acknowledge support from the JSPS KAKENHI (Grant Numbers 21H05233 and 23H02052), the CREST (JPMJCR24A5), JST and World Premier International Research Center Initiative (WPI), MEXT, Japan. V.F. and S.S. acknowledge support from EPSRC grant EP/V007033/1, British Council and International Science Partnerships Fund Grant 1185409051 for Research Collaboration between UK and Japan.

# *Supplementary Information for*

# Built-in Bernal gap in large-angle-twisted monolayer-bilayer graphene

Alex Boschi[1,2], Zewdu M. Gebeyehu[1,2], Sergey Slizovskiy[3,4,5], Vaidotas Mišeikis[1,2], Stiven Forti[1,2], Antonio Rossi[1,2], Kenji Watanabe[6], Takashi Taniguchi[7], Fabio Beltram[8], Vladimir I. Fal'ko[3,4,5], Camilla Coletti[1,2,*], Sergio Pezzini[8,**]

[1]*Center for Nanotechnology Innovation @NEST, Istituto Italiano di Tecnologia, Piazza San Silvestro 12, I-56127 Pisa, Italy*
[2]*Graphene Labs, Istituto Italiano di Tecnologia, Via Morego 30, 16163 Genova, Italy*
[3]*National Graphene Institute, The University of Manchester, Manchester, M13 9PL, UK*
[4]*School of Physics & Astronomy, The University of Manchester, Oxford Rd., Manchester, M13 9PL, UK*
[5]*Henry Royce Institute for Advanced Materials, Manchester, M13 9PL, UK*
[6]*Research Center for Electronic and Optical Materials, National Institute for Materials Science, 1-1 Namiki, Tsukuba, 305-0044, Japan*
[7]*Research Center for Materials Nanoarchitectonics, National Institute for Materials Science, 1-1 Namiki, Tsukuba, 305-0044, Japan*
[8]*NEST, Istituto Nanoscienze-CNR and Scuola Normale Superiore, Piazza San Silvestro 12, I-56127 Pisa, Italy*

### Supplementary Note 1. Optical microscopy images of the TMBG devices

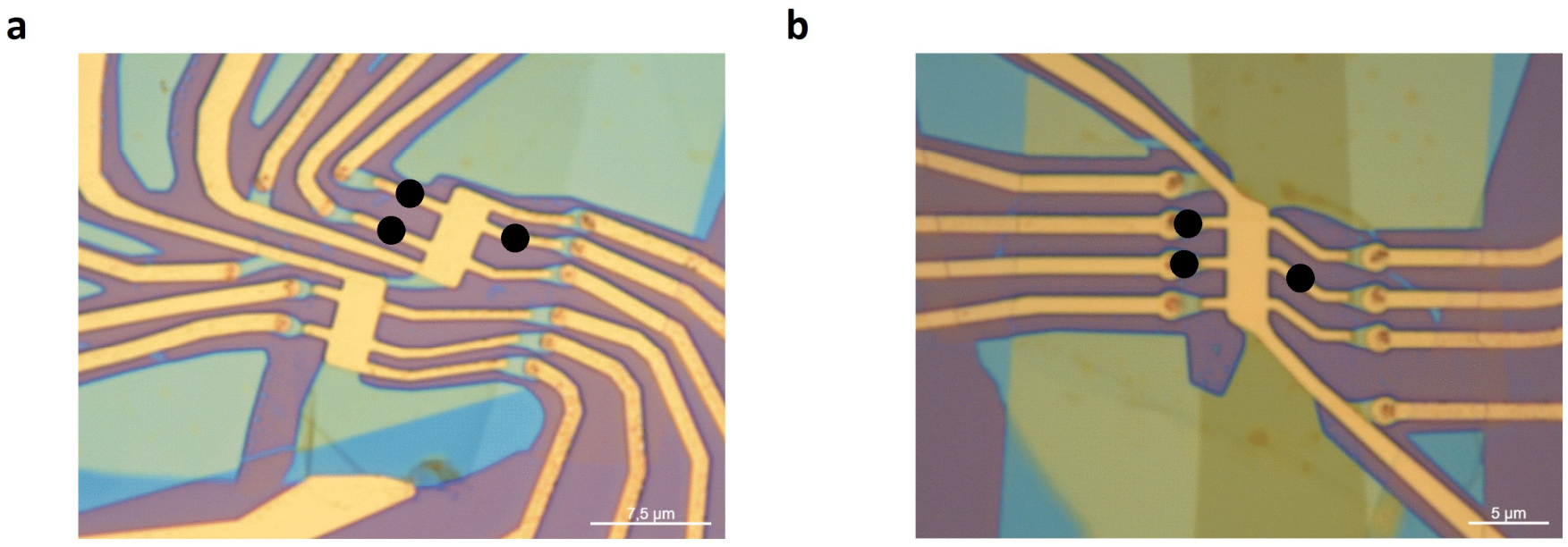

**Figure S1.** *Optical microscopy images of devices D1 (a) and D2 (b). The scale bars are 7.5 μm and 5 μm in (a) and (b), respectively. The voltage probes used for the measurements presented in the Main Text (D1) and Supporting Information (D2) are indicated by black dots.*

### Supplementary Note 2. Raman spectra of CVD-grown graphene stacks on $SiO_2$

We use Raman spectroscopy to preliminarily characterize the CVD-grown TMBG crystals after transfer to $SiO_2$/Si. We find that TMBG shows a distinctive Raman signature in the 2D mode, which can be decomposed into a MLG-like and BLG-like component. For comparison, in Figure S2 we show full spectra acquired on TMBG, TBG, BLG and MLG crystals from the same growth batch under the same experimental conditions.

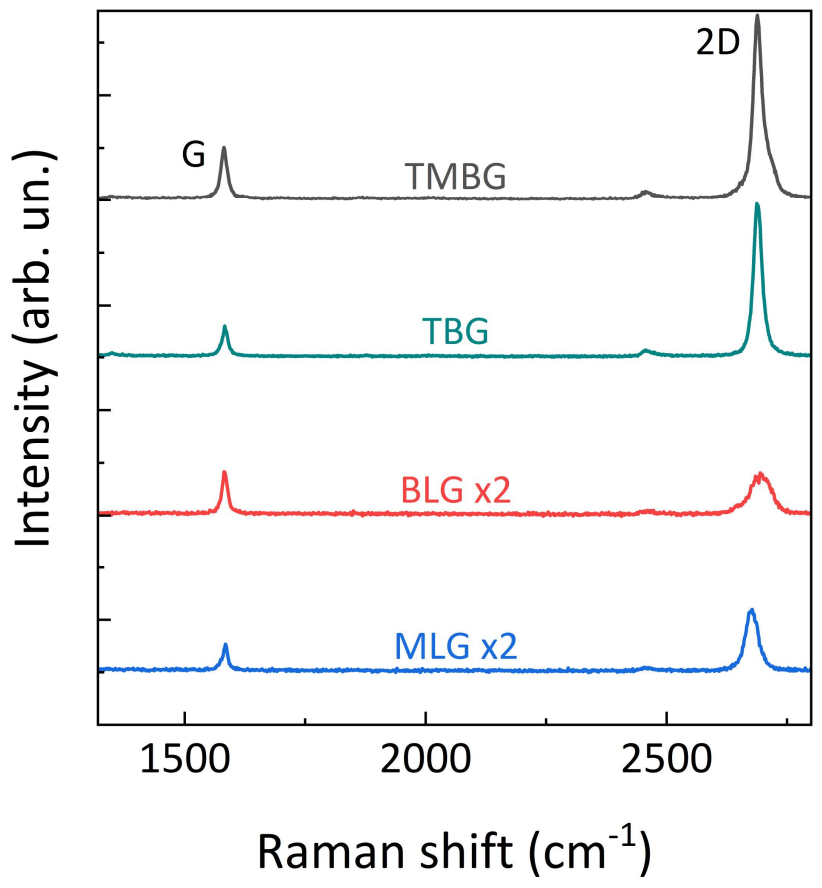

**Figure S2.** *Raman spectra of CVD-grown TMBG (dark grey), TBG (dark cyan), BLG (red) and MLG (blue) crystals transferred on $SiO_2$/Si. The spectra are vertically shift for clarity. The BLG and MLG spectra are multiplied by a factor two to facilitate comparison.*

## Supplementary Note 3. Theoretical calculation of gated TMBG with proximity-induced energy shifts

Due to a large twist angle between monolayer and bilayer components of TMBG, its electronic spectrum is a superposition of decoupled monolayer (MLG) and AB-stacked bilayer (BLG) spectra, each described in terms of effective k.p Hamiltonians in the vicinity of the respective K-points,

$$H_{MLG} = \begin{pmatrix} U_3 + \delta & \tilde{v}\,\tilde{\pi}^\dagger \\ \tilde{v}\,\tilde{\pi} & U_3 + \delta \end{pmatrix}$$

$$H_{BLG} = \begin{pmatrix} U_2 & v\,\pi^\dagger & -v_4\,\pi^\dagger & v_3\pi \\ v\,\pi & U_2 + \Delta_{AB} & \gamma_1 & -v_4\pi^\dagger \\ -v_4\pi & \gamma_1 & U_1 + \delta + \Delta_{AB} & v\,\pi^\dagger \\ v_3\pi^\dagger & -v_4\pi & v\,\pi & U_1 + \delta \end{pmatrix}$$

$$U_2 = U_1 + e\,D_1 c_1 - e\,l\frac{D_0 + D_1}{4} - e\,l\frac{D_1 + D_2}{4}$$

$$U_3 = U_2 + e\,D_2 c_2 - e\,l\frac{D_1 + D_2}{4} - e\,l\frac{D_2 + D_3}{4}$$

$$\frac{e^2 n_i}{\epsilon_0} = e\,D_{i-1} - e\,D_i \;\;,$$

where we implement self-consistent on-layer electrostatic potentials $U_{1,2,3}$, as marked on Fig S3a. The Hamiltonian $H_{BLG}$is written in the sublattice basis $(B_2, A_2, B_1, A_1,)$. Here, the integral TMBG charge density and the displacement field (controlled by the external gates) are introduced as

$$n_{tot} \equiv \frac{1}{e}\left(C_{tg}V_{tg} + C_{bg}V_{bg}\right) = n_1 + n_2 + n_3, \quad D \equiv \frac{\epsilon_0}{2}\left(C_{bg}V_{bg} - C_{tg}V_{tg}\right) = \frac{D_0 + D_3}{2}.$$

The on-layer electrostatic potentials $U_i$ are related via displacement fields $D_i$ (see Fig S3a), screened by out-of-plane polarization of graphene sheets [1]. The screening can be described in terms of a "dielectric thickness" of graphene $l = \frac{\alpha}{A_{cell}} \approx \frac{10.8\ \text{Å}^3}{5.2\ \text{Å}^2} = 2.06\ \text{Å}$ , where $\alpha \approx 10.8\ \text{Å}^3$ is an out-of-plane polarizability for a unit cell of graphene and we choose the interplane distances $c_1 = 3.35\ \text{Å}$, $c_2 = 3.44\ \text{Å}$ as known for Bernal and turbostratic graphites respectively. We note that the displacement field experienced by BLG is given by $\frac{D_0+D_2}{2}$. The difference with respect to $D$ stems from the electric field of MLG, which is proportional to MLG doping. In the conditions of the reported experiment, however, this contribution is very small, since MLG doping is much smaller than BLG doping.

In the single-particle part of MLG and BLG Hamiltonians, operators $\pi$ ( $\tilde{\pi}$ ) are complex momenta, $\mp i\,\hbar\,\partial_x + \hbar\,\partial_y - i\,e\,B\,x$, counted from $K_\pm$ points of BLG (MLG); $\tilde{v} = 1.15 \cdot 10^6$ m/s and $v = 1.02 \cdot 10^6$ m/s are Dirac velocities in MLG and BLG, which are slightly different due to renormalization by Coulomb interaction in MLG [2]; parameters $v_i = \frac{\sqrt{3}}{2} a\,\gamma_i$ are determined by Slonczewski-Weiss-McClure skew-hopping energies, $\gamma_3$ =0.38 eV and $\gamma_4$ =0.2 eV, and $a = 0.246$ nm is graphene's lattice constant. To mention, $\gamma_4$ contributes to the electron-hole asymmetry in the BLG, producing a slightly larger effective mass for holes, as compared to electrons. As usual, $\gamma_1$ =0.39 eV is energy of the closest-neighbour (dimer $A_2 - B_1$) interlayer hopping, and $\Delta_{AB} = 0.02$ eV is a dimer—non-dimer energy difference in BLG. In the context of the present study, we also take into account interlayer-proximity-induced energy shifts $\delta$ on layers 1 and 3 which interface the encapsulating hBN.

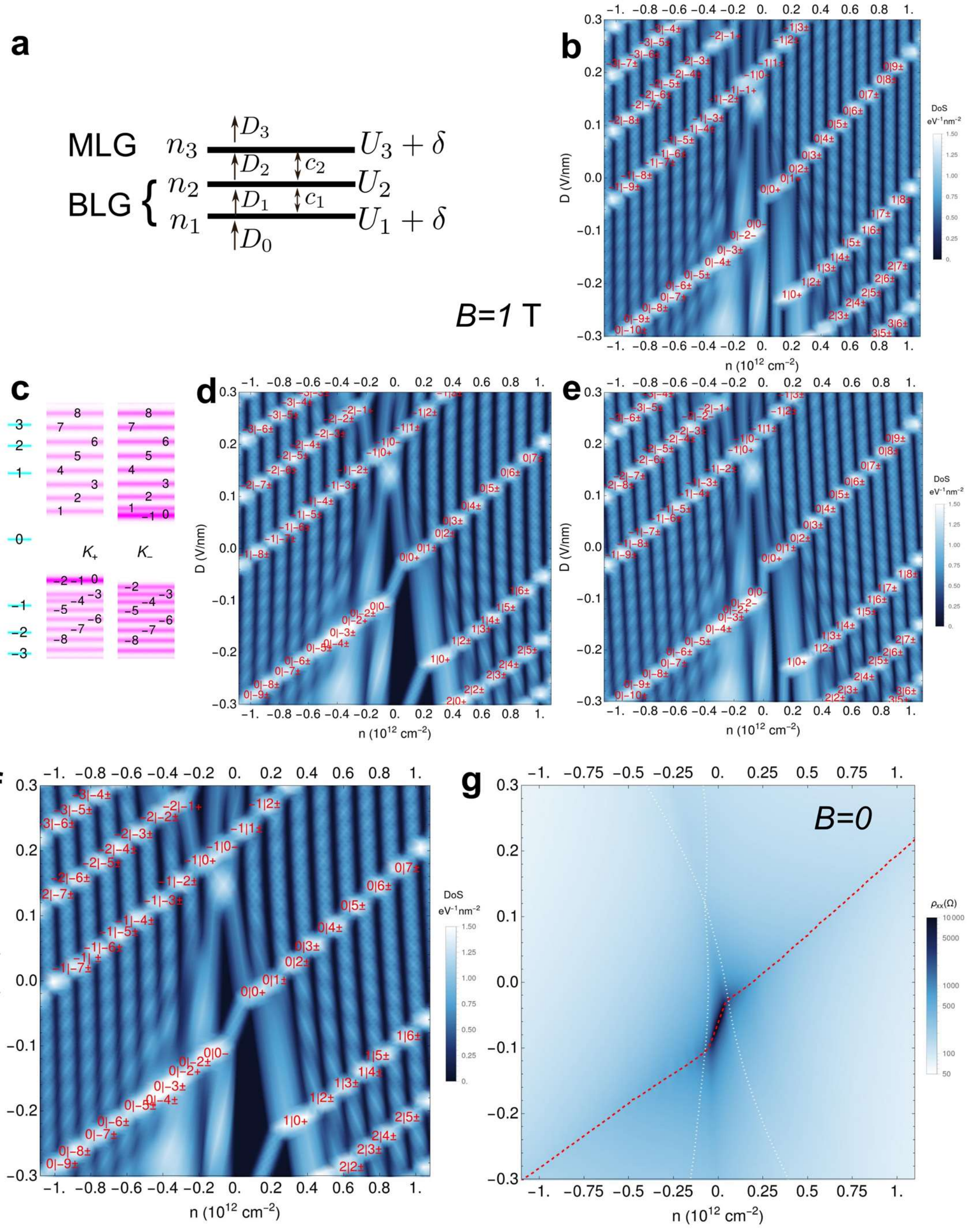

**Figure S3.** (a) *Notations used in self-consistent calculations. Energy is counted from the Fermi level.* (b,d,e,f) *Calculated density of states for MLG and BLG parts of the stack, B = 1 T, numbers of mutually pinned MLG and BLG LLs are shown at level-crossings. Level-numbering scheme for MLG and BLG is shown in* (c). *In* (b) *we assume no localized states, in* (d) *we assume* $\rho = 0.08\ \mathrm{eV^{-1}nm^{-2}}$*, in* (e) $\tilde{\rho} = 0.01\ \mathrm{eV^{-1}nm^{-2}}$ *and in* (f) *we assume localized states on both MLG and BLG.* (g) *Calculated resistivity of TMBG for B = 0, assuming a fixed scattering time* $\tau = 10^{-12}\ s$*. Red dashed line indicates the position of MLG Dirac point and white dotted lines correspond to valence and conduction band edges of BLG.*

In a magnetic field, $B$, the Peierls substitution makes operator $\pi = \mp i\,\hbar\,\partial_x + \hbar\,\partial_y - i\,e\,B\,x$ identical to a lowering/raising operator acting in the basis of magnetic oscillator states $\phi_j$. Taking into account the difference between $K_\pm$ valleys, this determines $\pi\,\phi_j = -\frac{i\hbar\sqrt{2\,j}}{l_B}\,\phi_{j-1}$ for $K_+$ valley and $\pi\,\phi_j = \frac{i\hbar\sqrt{2\,(j+1)}}{l_B}\,\phi_{j+1}$ for $K_-$ valley, where $l_B = \sqrt{\frac{\hbar}{eB}}$. For the MLG LL spectrum we use its established analytical form, $\tilde{E}_n = U_3 + \delta + \mathrm{sign}(n)\,\sqrt{2|n|}\hbar\frac{v}{l_B}$ (shifted by the on-layer potential) and evaluate the electron density on it, taking into account the valley degeneracy of MLG LLs,

$$n_3 = \frac{2\,e\,B}{\pi\,\hbar}\sum_{n=-N}^{N}\int\frac{\tau\,dE}{2\,\hbar}\mathrm{sech}^2\frac{(E-\tilde{E}_n)\tau}{\hbar}\left[n_F(E)-\frac{1}{2}\right]$$

where $N$ is a cutoff on the number of LLs included, $n_F$ is Fermi function and a -1/2 factor compensates for the Fermi sea density. Also, we introduce LL broadening by a factor $\frac{\tau}{2\,\hbar}\mathrm{sech}^2\frac{(E-\tilde{E}_n)\tau}{\hbar}$, where we choose $\hbar\tau^{-1}$= 0.7 meV as a representative value for computing density of states maps. As the BLG Hamiltonian is more complicated, the BLG LL spectrum is computed numerically in the basis of 400 Landau oscillator states on sublattices (101/100/100/99 oscillator states on $B_2/A_2/B_1/A_1$ sublattices in valley $K_+$ and 99/100/100/101 in $K_-$ valley respectively), simultaneously with the self-consistent calculation of on-layer potentials and charge densities on all layers. This includes finding *4N* eigenvectors $\Psi^{i\pm}_{m,A/B}$, $i_\pm = -N,\dots,N-1$, where $m = 1{,}2$ is a BLG layer number as marked on Fig S3a and $\pm$ identifies $K_\pm$ valleys, as marked in sketch S3c. We note that such counting assumes that -1 and 0 LLs originate from 0-energy LLs in gapless BLG, [3] which valley degeneracy is lifted by the interplay of interlayer asymmetry and time-reversal breaking by a magnetic field. The electron density on BLG layers is computed as

$$n_m = \sum_{K_\pm}\frac{2\,e\,B}{2\,\pi\,\hbar}\sum_{i=-N}^{N-1}\left[\left|\Psi^{i\pm}_{m,A}\right|^2 + \left|\Psi^{i\pm}_{m,B}\right|^2\right]\int\frac{\tau\,dE}{2\,\hbar}\mathrm{sech}^2\frac{(E-E^i_\pm)\tau}{\hbar}\left[n_F(E)-\frac{1}{2}\right].$$

The above-described self-consistent analysis of LL spectrum of TMBG produces density of states (DOS) maps shown in Fig. S3b. Here we highlight a specific effect of pinning of BLG LLs (vertical on the plot) to MLG LL (going diagonally) at level-crossings.

As the above calculations assume narrow LLs, chemical potential passes all inter-LL gap instantly upon changing carrier density, so that BLG gap does not show up on the *n-D* map in Fig.S3b. To account for the experimentally observed finite range of doping where chemical potential in TMBG is inside a BLG gap, we incorporate in the model a constant DOS of localized states, $\rho$ and $\tilde{\rho}$ for MLG and BLG, respectively. Those originate from short-range defects and hybridization of graphene states with occasional adatoms, with the energies of the localized states systematically pushed towards Dirac point[4], resulting in additional states outside the LL basis at low energies as they are provided by the part of Hilber space originating from distant areas of graphene Brillouin zone. To take into account the contribution of those strongly-localized states, we amend the on-layer densities as,

$$n_m = \frac{-1}{2}\,\rho\,\frac{U_1+U_2}{2} + \sum_{K_\pm}\frac{2\,e\,B}{2\,\pi\,\hbar}\sum_{i=-N}^{N-1}\left[\left|\Psi^i_{m,A,\pm}\right|^2 + \left|\Psi^i_{m,B,\pm}\right|^2\right]\int\frac{\tau\,dE}{2\,\hbar}\,\mathrm{sech}^2\frac{(E-E^i_\pm)\tau}{\hbar}\left[n_F(E)-\frac{1}{2}\right],$$

$$n_3 = -(U_3+\delta)\tilde{\rho} + \frac{2\,e\,B}{\pi\,\hbar}\sum_{n=-N}^{N}\int\frac{\tau\,dE}{2\,\hbar}\,\mathrm{sech}^2\frac{\left(E-\tilde{E}_n\right)\tau}{\hbar}\left[n_F(E)-\frac{1}{2}\right].$$

The results of numerical simulations taking into account such disorder either only in the BLG or only in the MLG are shown in Fig. S3d and S3e, respectively (for $\rho = 8\cdot 10^{12}\ \mathrm{eV^{-1}cm^{-2}}$ and $\tilde{\rho} = 10^{12}\ \mathrm{eV^{-1}cm^{-2}}$). Then we note that the experimentally obtained QHE maps are described well by choosing $\rho = 8\cdot 10^{12}\ \mathrm{eV^{-1}cm^{-2}}$ and $\tilde{\rho} = 10^{12}\ \mathrm{eV^{-1}cm^{-2}}$, which is plotted in Fig. S3f.

To estimate conductivity at $B$ = 0, we assumed the same scattering time $\tau$ for carriers in MLG and BLG and computed the conductivity as $\sigma = \frac{-4e^2\tau}{(2\,\pi\,\hbar)^2}\left[\langle v_x^2\rangle\,DoS + \langle\tilde{v}_x^2\rangle\,\widetilde{DoS}\right]$ for BLG and MLG parts, where localized states were excluded from the DOS in this expression. The electronic density has been computed by accounting for the same electron lifetime $\tau$ as used above, for MLG we get,

$$n_3 = -(U_3+\delta)\tilde{\rho} + \frac{1}{\pi\hbar^2 v^2}\int\frac{\tau\,dE}{2\,\hbar}\,E^2\mathrm{sign}(E)\,\mathrm{sech}^2\frac{(E+U_3+\delta)\tau}{\hbar}$$

$$= -(U_3+\delta)\tilde{\rho} + \frac{1}{2\pi v^2\tau^2}\left[\mathrm{Li}_2\left(-e^{2(U_3+\delta)\frac{\tau}{\hbar}}\right) - \mathrm{Li}_2\left(-e^{-2(U_3+\delta)\frac{\tau}{\hbar}}\right)\right],$$

where $\mathrm{Li}_2$ is a polylogarithm function. For BLG part, we compute the spectrum and eigenfunctions and compute the density as

$$n_m = \frac{-1}{2}\,\rho\,\frac{U_1+U_2}{2} + \int\frac{d^2k}{\pi^2}\sum_{\beta=1}^{4}\left[\left|\Psi^\beta_{m,A}(k)\right|^2 + \left|\Psi^\beta_{m,B}(k)\right|^2\right]\int\frac{\tau\,dE}{2\,\hbar}\,\mathrm{sech}^2\frac{(E-E^\beta(k))\tau}{\hbar}\left[n_F(E)-\frac{1}{2}\right]$$

The resulting resistivity $\rho = 1/\sigma$ map is shown in Fig. S3g (and in Figure 2c inset in the Main Text). It displays distinct maximum values in the range where Dirac point in MLG crosses the BLG gap. Outside that interval of densities and displacement field one still can trace the conditions when the Dirac point in MLG is at the Fermi level, which is highlighted by red dashed line. We also mark points on *n-D* map, where the Fermi level in the system crosses conduction and valence band edges of gapped BLG (white dots), which coincide with some weaker features on the resistivity map in Fig S3g.

**Supplementary Note 4. Transport data from device D2**

We repeat the transport measurements discussed in the Main Text on a second TMBG device, D2 (see optical microscopy image in Figure S1b). In Figure S4a we show the zero-field resistivity (at $T$ = 2.7 K) as a function of $n_{tot}$ and $D$, which reproduces the key features presented for D1 in Figure 2b. Minor differences in the resistivity values suggest a slightly higher carrier mobility for this sample. Figure S4b shows resistivity curves at different $D$, highlighting the asymmetry with respect to the displacement field. In Figure S4c we apply a small perpendicular magnetic field ($B$ = 0.25 T, while $B$ = 1 T is used for D1 in Figure 2e), which induces Landau quantization in the MLG subsystem, enhancing the visibility of the MLG CNP and the BLG gap region (dark blue area). $D$ and $n_{tot}$ are computed considering a slight thickness difference between bottom and top hBN flakes (~36 nm and ~30 nm), corresponding to $C_{bg}$ = 7.38 x $10^{-8}$ $Fcm^{-2}$and $C_{tg}$ = 8.85 x $10^{-8}$ $Fcm^{-2}$, respectively.

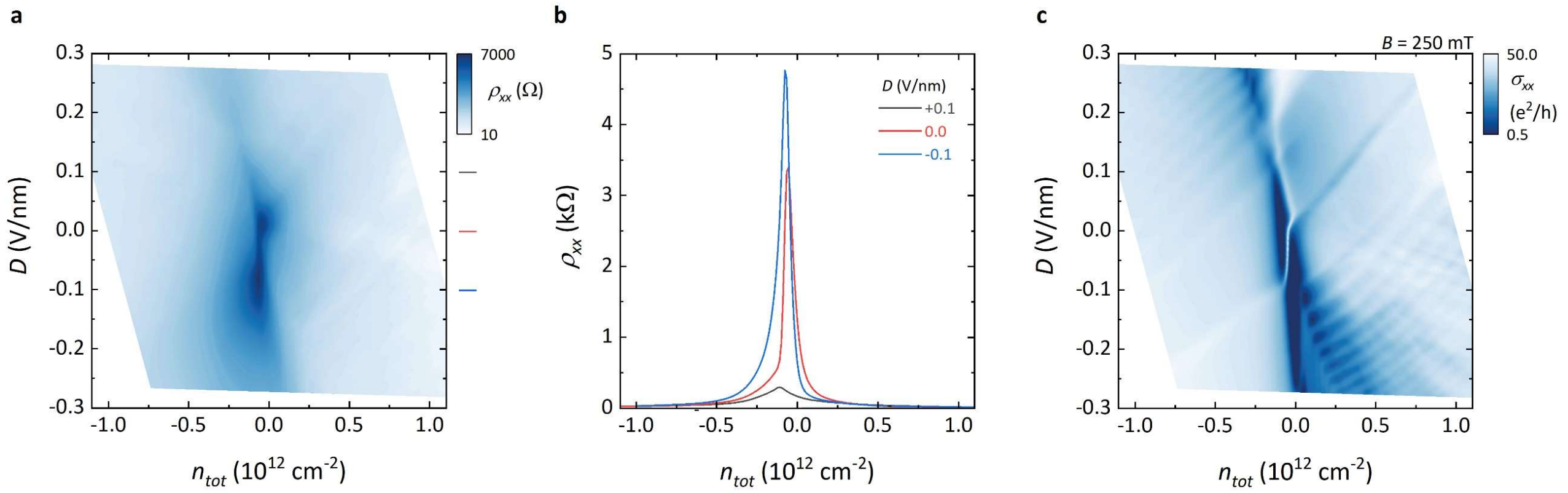

**Figure S4.** *(a) Resistivity as a function of carrier density and displacement field for sample D2, acquired within the same gate ranges used in Figure 2b for D1. (b) Resistivity of D2 as a function of carrier density at selected displacement fields, as marked in panel (a). (c) Longitudinal conductivity at B = 0.25 T, as a function of carrier density and displacement field (same ranges as in (a)). A logarithmic color scale is used in panels (a) and (c). Data acquired at T = 2.7 K.*

**Supplementary Note 5. Transport data from 30°-twisted MLG-BLG-MLG**

The structural asymmetry critically influences the electronic properties of TMBG. To further support this central finding, we fabricate a third device (D3), in which the same subsystems (MLG and BLG) are arranged in a symmetric 30°-twisted configuration. The structure of the device is shown in Figure S5a. In this sample, BLG (red and orange layers) is encapsulated by two rotated MLG on both sides. The stack is obtained via the grow-and-stack method introduced in Ref. 5, in which CVD-grown single-crystals undergo an angle-controlled hBN-mediated stacking. The 2D Raman mode of this system (see Figure S5b) shows strong similarities with that of TMBG, with convoluted MLG-like and BLG-like components. The stronger sharp Lorentzian peak is consistent with the presence of two MLGs in the stack. Figure S5c shows the low-temperature transport response as a function of the top and bottom gate potentials (note that in this device the bottom gate is provided by the underlying p-doped Si wafer with 300 nm $SiO_2$, which requires larger applied voltages with respect to the top gate acting through ~30 nm hBN). The device shows a clear symmetry with respect to the displacement field. The features along the main diagonal are associated with the BLG subsystem, which acquires a band gap only at finite *D*. Two square-root-like features (see guides-to-the-eye in Figure S5c) mark the CNPs form the two MLGs, which position in the stack can be easily inferred from their gate dependence (the top one being more sensitive to the top gate, vice-versa for the bottom one). Different gate-controlled band alignments between the three subsystems are also indicated.

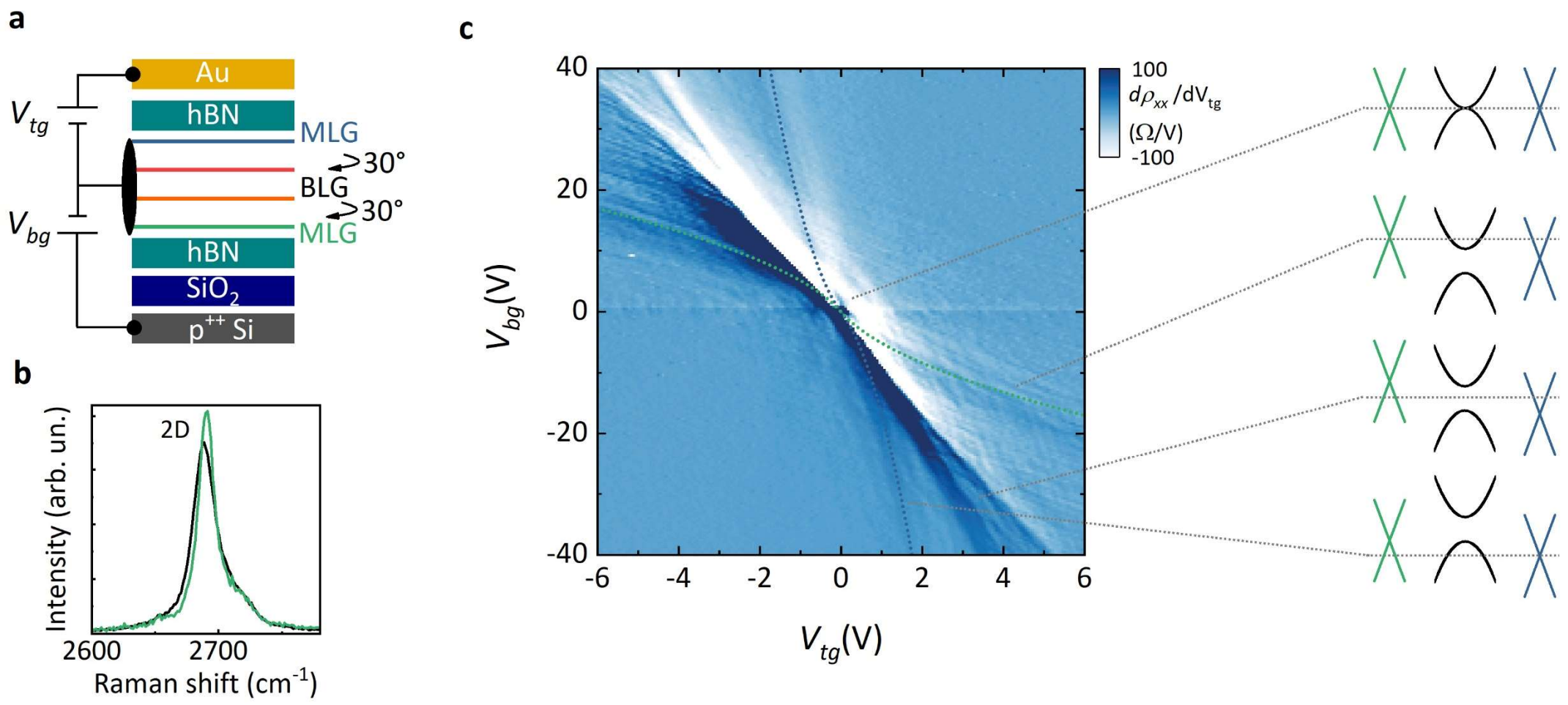

**Figure S5.** *(a) Side-view sketch of device D3. (b) 2D Raman mode measured on MLG-BLG-MLG (green curve). The black curve is from TMBG (same data shown in Figure 1e). (c) First derivative of the longitudinal resistivity of D3 as a function of the gate potentials, measured at T = 4.2 K. The blue and green dotted lines are guides-to-to-eye helping to identify the CNPs from the two MLG subsystems. Examples of gate-tuned band alignments are sketched on the righthand side.*

## Supplementary Note 6. Thermal activation measured in sample D2

In Figure S6, we present the thermal activation measurements of the BLG band gap performed on TMBG device D2. Figure S6a shows the resistivity as a function of the carrier density at different temperatures (with the displacement field *D* = -0.09 V/nm setting the MLG CNP within the BLG gap). We observe a strong thermal activation of the resistivity peak, consistent with sample D1. Figure S6b shows the Arrhenius plot for the resistivity maximum (blue dots), with the corresponding thermal-activation fit (continuous line). The activation gap and its dependence on the displacement field (Figure S6b inset) are compatible with our findings for sample D1.

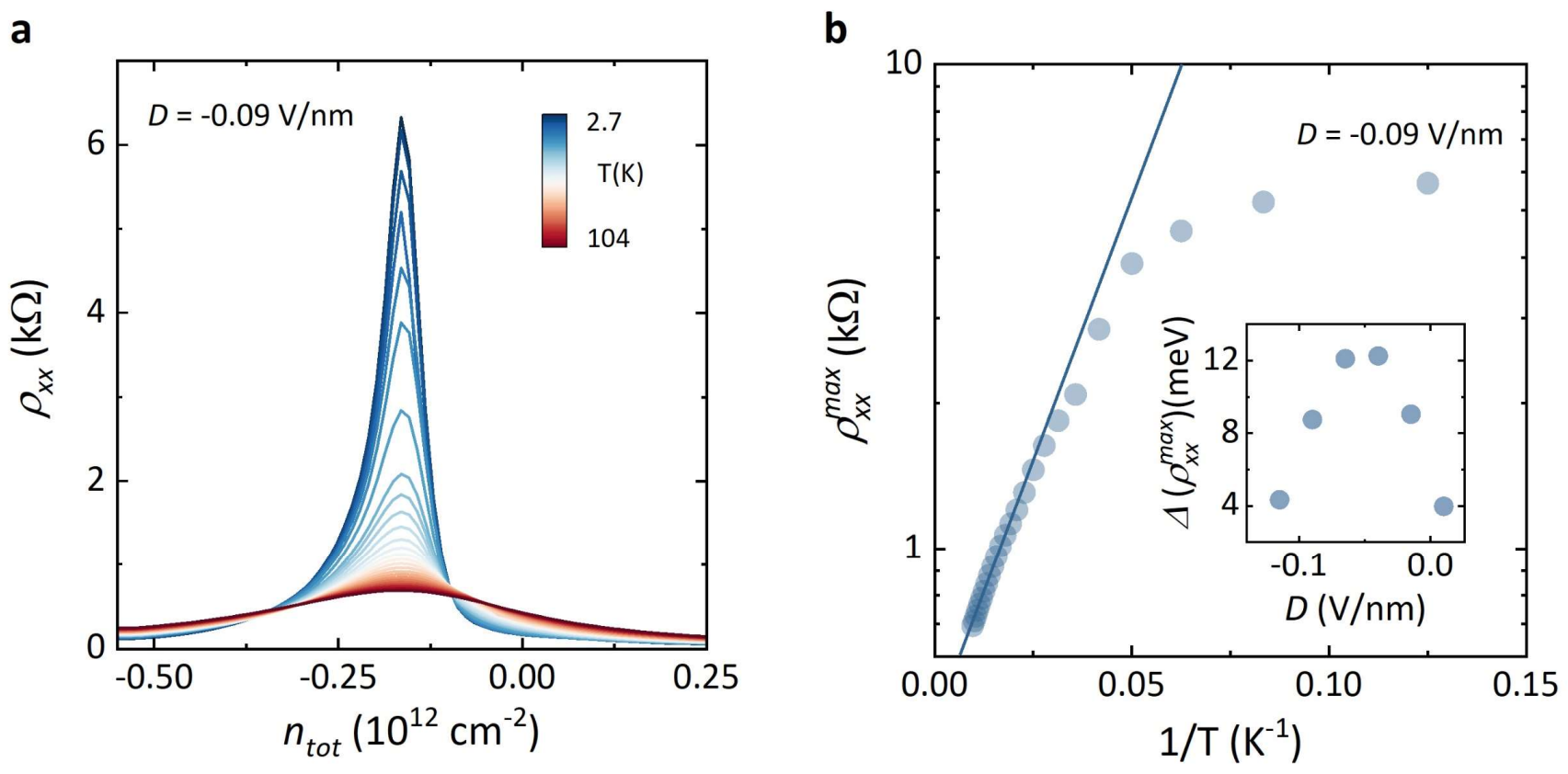

**Figure S6.** *(a) Longitudinal resistivity as a function of the carrier density, measured on device D2 at D = -0.09 V/nm for 2.7 K < T < 104 K. (b) Arrhenius plot of the resistivity maxima from data in (a) (blue dots). The blue continuous line is a fit to* $\rho_{xx}^{max} \propto \exp[\Delta/(2k_BT)]$. *Inset: activation gaps obtained from the Arrhenius fit at different displacement fields.*

## Supplementary References